# Distance-based Chatterjee correlation: a new generalized robust measure of directed association for multivariate real and complex-valued data

Roberto D. Pascual-Marqui[1], Kieko Kochi[1], Toshihiko Kinoshita[2]

1: The KEY Institute for Brain-Mind Research; Department of Psychiatry; University of Zurich, Switzerland
2: Department of Neuropsychiatry, Kansai Medical University, Osaka, Japan

Corresponding author: RD Pascual-Marqui
robertod.pascual-marqui@uzh.ch ; https://www.uzh.ch/keyinst
https://scholar.google.com/citations?user=DDqjOkUAAAAJ

## 1. Abstract

Building upon the Chatterjee correlation (2021: J. Am. Stat. Assoc. 116, 2009–2022) for two real-valued variables, this study introduces a generalized measure of directed association between two vector variables, real or complex-valued, and of possibly different dimensions. The new measure is denoted as the "distance-based Chatterjee correlation", owing to the use here of the "distance transformed data" defined in Székely et al (2007: Ann. Statist. 35, 2769-2794) for the distance correlation. A main property of the new measure, inherited from the original Chatterjee correlation, is its predictive and asymmetric nature: it measures how well one variable can be predicted by the other, asymmetrically. This allows for inferring the causal direction of the association, by using the method of Blöbaum et al (2019: PeerJ Comput. Sci. 1, e169). Since the original Chatterjee correlation is based on ranks, it is not available for complex variables, nor for general multivariate data. The novelty of our work is the extension to multivariate real and complex-valued pairs of vectors, offering a robust measure of directed association in a completely non-parametric setting. Informally, the intuitive assumption used here is that distance correlation is mathematically equivalent to Pearson's correlation when applied to 'distance transformed' data. The next logical step is to compute Chatterjee's correlation on the same 'distance transformed' data, thereby extending the analysis to multivariate vectors of real and complex valued data. As a bonus, the new measure here is robust to outliers, which is not true for the distance correlation of Székely et al. Additionally, this approach allows for inference regarding the causal direction of the association between the variables.

## 2. The Chatterjee correlation

The non-parametric correlation coefficient introduced by Chatterjee (2021) for two real-valued variables has many important properties, three of which are of main interest here: it takes the value zero if and only if the variables are independent; it is not symmetric in the two variables; and it measures how well one variable can be predicted by the other variable.

Consider the real-valued paired data $(x_1, y_1), ..., (x_N, y_N)$, and define the re-ordered pairs $(x_{(1)}, y_{(1)}), ..., (x_{(N)}, y_{(N)})$ such that $x_{(1)} \leq ... \leq x_{(N)}$. Let $r_i = rank(y_{(i)})$ denote the rank of $y_{(i)}$. Then Chatterjee's correlation (2021) for "y as a function of x" is:

**Eq. 1** $$Ch[y = f(x)] = 1 - \frac{3\sum_{k=1}^{N-1}|r_{k+1} - r_k|}{N^2 - 1}$$

and it is to be interpreted as a measure of how well "*y* is predicted by *x*".

Chatterjee (2021) also considers the case of data with ties (see the non-numbered equation between equations 1 and 2 therein).

Due to the use of ranks, there is no trivial extension of the Chatterjee correlation for complex-valued data, nor for multivariate vectors.

## 3. The distance transform for data

Consider paired multivariate data $\mathbf{X}_i \in \mathbb{C}^{p \times 1}$ and $\mathbf{Y}_i \in \mathbb{C}^{q \times 1}$ for $i = 1...N$. Note that this includes the case for real-valued vectors as well.

The distance transform used here, which is based on Székely et al (2007), takes data such as $\mathbf{X}_i \in \mathbb{C}^{p \times 1}$ for $i = 1...N$, and converts it to real-valued scalars $D_k^x \in \mathbb{R}$ for $k = 1...M$, with:

**Eq. 2** $M = N(N+1)/2$

The distance transform is defined as follows. For **X**, first define the matrix $\mathbf{a} \in \mathbb{R}^{N \times N}$ with elements:

**Eq. 3** $a_{ij} = \sqrt{(\mathbf{X}_i - \mathbf{X}_j)^*(\mathbf{X}_i - \mathbf{X}_j)}$ ; for $i, j = 1...N$

where the superscript "*" denotes, in general, complex conjugate and transpose. Next, define the double centered matrix $\mathbf{A} \in \mathbb{R}^{N \times N}$:

**Eq. 4** $\mathbf{A} = \mathbf{HaH}$

where $\mathbf{H} \in \mathbb{R}^{N \times N}$ is the centering matrix:

**Eq. 5** $\mathbf{H} = \mathbf{I} - \frac{1}{N}\mathbf{11}^T$

and where **I** is the identity matrix and **1** is a vector of ones. Finally, the elements in the upper triangle of **A**, including the diagonal, define the distance transformed data for **X**:

**Eq. 6** $(D_k^x \text{ for } k = 1..M) \Leftrightarrow (D_1^x, D_2^x, ..., D_M^x) \Leftrightarrow (A_{11}, A_{12}, ..., A_{1N}, A_{22}, A_{23}, ..., A_{2N}, A_{33}, A_{34}, ..., A_{3N}, ..., A_{N,N})$

where *M* is given by Eq. 2.

Similarly for **Y**, first define $\mathbf{b} \in \mathbb{R}^{N \times N}$ with elements:

**Eq. 7** $\quad b_{ij} = \sqrt{(\mathbf{Y}_i - \mathbf{Y}_j)^* (\mathbf{Y}_i - \mathbf{Y}_j)}$

Next, define:

**Eq. 8** $\quad \mathbf{B} = \mathbf{HbH}$

Finally, the elements in the upper triangle of **B** including the diagonal define the distance transformed data for **Y**:

**Eq. 9** $\quad \left(D_k^y \text{ for } k=1..M\right) \Leftrightarrow \left(D_1^y, D_2^y, ..., D_M^y\right) \Leftrightarrow \left(B_{11}, B_{12}, ..., B_{1N},\ B_{22}, B_{23}, ..., B_{2N},\ B_{33}, B_{34}, ..., B_{3N},\ ..,\ B_{N,N}\right)$

as real-valued scalars $D_k^y \in \mathbb{R}$ for $k = 1...M$, where $M$ is given by Eq. 2.

## 4. The new distance-based Chatterjee correlation

The non-parametric correlation coefficient introduced by Chatterjee (2021) can now be extended to multivariate real and complex-valued vectors. Let the real-valued pairs $\left(D_1^x, D_1^y\right), ..., \left(D_M^x, D_M^y\right)$ denote the data transforms for $\left(\mathbf{X} \in \mathbb{C}^{p \times 1}, \mathbf{Y} \in \mathbb{C}^{q \times 1}\right)$, as defined above. And let $\left(D_{(1)}^x, D_{(1)}^y\right), ..., \left(D_{(M)}^x, D_{(M)}^y\right)$ denote the reordered pairs such that $D_{(1)}^x \leq ... \leq D_{(M)}^x$. Let $\beta_i = rank\left(D_{(i)}^y\right)$ denote the rank of $D_{(i)}^y$. Then the new distance-based Chatterjee correlation, introduced here, for "**Y** as a function of **X**" is:

**Eq. 10** $\quad dCh\left[\mathbf{Y} = \mathbf{F}(\mathbf{X})\right] = 1 - \dfrac{3 \sum_{k=1}^{M-1} |\beta_{k+1} - \beta_k|}{M^2 - 1}$

and it is to be interpreted as a measure of how well "**Y** is predicted by **X**".

At the risk of over-redundancy, we explicitly include the definition for the distance-based Chatterjee correlation for "**X** as a function of **Y**", which is trivially calculated by exchanging the roles of **X** and **Y** above. Let the real-valued pairs $\left(D_1^y, D_1^x\right), ..., \left(D_M^y, D_M^x\right)$ denote the data transforms for $\left(\mathbf{Y} \in \mathbb{C}^{q \times 1}, \mathbf{X} \in \mathbb{C}^{p \times 1}\right)$, as defined above. And let $\left(D_{(1)}^y, D_{(1)}^x\right), ..., \left(D_{(M)}^y, D_{(M)}^x\right)$ denote the reordered pairs such that $D_{(1)}^y \leq ... \leq D_{(M)}^y$. Let $\alpha_i = rank\left(D_{(i)}^x\right)$ denote the rank of $D_{(i)}^x$. Then the new distance-based Chatterjee correlation, introduced here, for "**X** as a function of **Y**" is:

**Eq. 11** $\quad dCh\left[\mathbf{X} = \mathbf{G}(\mathbf{Y})\right] = 1 - \dfrac{3 \sum_{k=1}^{M-1} |\alpha_{k+1} - \alpha_k|}{M^2 - 1}$

and it is to be interpreted as a measure of how well "**X** is predicted by **Y**".

Only continuous data without ties is treated here. See Chatterjee (2021) for the case of data with ties (see the non-numbered equation between equations 1 and 2 therein).

## 5. Causal direction inference

The "regression error based causal inference" (RECI) algorithm of Blöbaum et al (2019) can be directly applied here for inferring the causal direction of the association between two vectors **X** and **Y**. Instead of regression error, "goodness of prediction" will be used here. Recall that the distance-based Chatterjee correlation is an asymmetric quantitative measure of how well one variable predicts the other, which is not the case for most other measures of correlation, such as e.g. Pearson, Spearman, and distance correlation. Although see Junker et al (2021) for an asymmetric measure and its use in estimating directed dependence limited to the case of two univariate variables.

Define the statistics:

**Eq. 12** $\Delta(x \to y) = dCh[\mathbf{Y} = \mathbf{F}(\mathbf{X})] - dCh_x[\mathbf{X} = \mathbf{G}(\mathbf{Y})]$

and:

**Eq. 13** $\Delta(y \to x) = dCh_x[\mathbf{X} = \mathbf{G}(\mathbf{Y})] - dCh[\mathbf{Y} = \mathbf{F}(\mathbf{X})] = -\Delta(x \to y)$

Then causal direction inference is defined as follows:
1. We say that **X** causes **Y** in the sense of RECI if $\Delta(x \to y) > 0$;
2. We say that **Y** causes **X** in the sense of RECI if $\Delta(y \to x) > 0$.

An important property here is that causation in the sense of RECI is not defined if the relation between **X** and **Y** is linear, since the prediction error is symmetric, i.e. $\Delta(x \to y) = \Delta(y \to x) = 0$.

## 6. Statistical inference for prediction and for causality based on synchronized random permutations

For the statistics defined in Eq. 10, Eq. 11, Eq. 12, and Eq. 13, the hypotheses of interest are:
1. $H_0 : dCh[\mathbf{Y} = \mathbf{F}(\mathbf{X})] = 0$, i.e. "**Y** is not predicted by **X**"; against $H_1 : dCh[\mathbf{Y} = \mathbf{F}(\mathbf{X})] > 0$, i.e. "**Y** is predicted by **X**".
2. $H_0 : dCh[\mathbf{X} = \mathbf{G}(\mathbf{Y})] = 0$, i.e. "**X** is not predicted by **Y**"; against $H_1 : dCh[\mathbf{X} = \mathbf{G}(\mathbf{Y})] > 0$, i.e. "**X** is predicted by **Y**".
3. $H_0 : \Delta(x \to y) = 0$, i.e. "**X** does not cause **Y**"; against $H_1 : \Delta(x \to y) > 0$, i.e. "**X** causes **Y**".
4. $H_0 : \Delta(y \to x) = 0$, i.e. "**Y** does not cause **X**"; against $H_1 : \Delta(y \to x) > 0$, i.e. "**Y** causes **X**".

The tests are carried out as follows. Given data $(\mathbf{X}_i, \mathbf{Y}_i)$ for $i = 1...N$, compute the distance transforms (Eq. 6 and Eq. 9), then compute the distance based Chatterjee correlations from Eq. 10 and Eq. 11, and the causal measures Eq. 12 and Eq. 13. Denote the values as:

**Eq. 14** $dCh_0[\mathbf{Y} = \mathbf{F}(\mathbf{X})]$, $dCh_0[\mathbf{X} = \mathbf{G}(\mathbf{Y})]$, $\Delta_0(x \to y)$, $\Delta_0(y \to x)$

Next, define a random permutation of indices $i = 1...N$ which will be used two times in:
- randomly reordering the $N$ vectors $\mathbf{Y}_i$ for computing $dCh_1^R[\mathbf{Y} = \mathbf{F}(\mathbf{X})]$ for Eq. 10,
- randomly reordering the $N$ vectors $\mathbf{X}_i$ for computing $dCh_1^R[\mathbf{X} = \mathbf{G}(\mathbf{Y})]$ for Eq. 11.

The use of the same permutation of indices is known as a synchronized permutation, see e.g. Winkler et al (2016).

The previous step produced one sample from the null hypotheses (no prediction, no causality), with values denoted as:

**Eq. 15** $dCh_k^R[\mathbf{Y} = \mathbf{F}(\mathbf{X})]$, $dCh_k^R[\mathbf{X} = \mathbf{G}(\mathbf{Y})]$, $\Delta_k^R(x \to y)$, $\Delta_k^R(y \to x)$

where the superscript "R" denotes random permutation, and the subscript "k" denotes the k-th synchronized permutation.

Next, the synchronized permutation step is repeated a large number of times $K$, thus $k=1...N$.

The set of all values for $k=1...K$ define the empirical probability distributions under the null hypotheses.

Let $s_0$ denote the value of any one of the statistics in Eq. 14, and let $s_1^R, s_2^R, ..., s_K^R$ denote the corresponding values that define its empirical probability distribution under the null hypothesis. Then the $p$-value is obtained as:

**Eq. 16** $$p = \frac{1}{K} \sum_{k=1}^{K} Ind\left(s_0 \leq s_k^R\right)$$

where $Ind(\bullet)$ is the indicator function that returns 1 if the argument is true and returns zero otherwise. A small value of $p$, say $p < 0.05$, favors the alternative hypothesis (significant prediction or significant causality).

Note that for two real variables, this same procedure applies to the original Chatterjee correlation (Eq. 1), by simply using the data and not its distance transformed version.

# 7. Statistical power analysis using synchronized random permutations on repeatedly generated toy data

Any proposal for a new measure of association must be compared, in terms of power, to other well-known and established measures. This is the main empirical result of interest, as shown in, e.g. the seminal papers of Székely et al (2007) and Chatterjee (2021), for the distance correlation and for the Chatterjee correlation.

In this study here, power analyses are carried out in a classical manner. The methodology is illustrated with a particular example that can be generalized to other settings and assumptions:
Step#0: Given a rule of association between **X** and **Y**, e.g. $y = \ln x^2 + \varepsilon$, with random $x$ uniformly distributed on $[-1,+1]$, independent of random $\varepsilon \sim N(0, \sigma_{\varepsilon\varepsilon})$. Given a sample size $N_{samp}$ (e.g. $N_{samp} = 100$). Given a noise level $\sigma_{\varepsilon\varepsilon}$. Given a significance level $\alpha$ (e.g. $\alpha=0.05$, or $\alpha=0.1$). Given the number of simulations $N_{sim}$ (e.g. $N_{sim} = 90$).
Step#1: For $i = 1...N_{sim}$ perform Step#1a and Step#1b:
　　Step#1a: Generate random data according to the rule and parameters defined in Step#0.
　　Step#1b: Estimate the $p$-value via synchronized random permutations (see above, Eq. 14 through Eq. 16) for anyone (or each one) of the hypotheses of interest for prediction and causality. Denote as $p_i$.
Step#2: Compute the power of the test under the model and the parameters specified in Step#0 as:

**Eq. 17** $$pow(model, \sigma_{\varepsilon\varepsilon}, N_{samp}, \alpha) = \frac{1}{N_{sim}} \sum_{i=1}^{N_{sim}} Ind(p_i < \alpha)$$

As shown, statistical power explicitly depends on the actual model of association, and on parameters such as sample size, noise level, and the significance level that was chosen.

Note that for two real variables, this same procedure applies to the original Chatterjee correlation (Eq. 1), by simply using the data and not its distance transformed version.

## 8. Example for bivariate simulated data: linear, w-shape, sinusoid, circular

Based on Chatterjee (2021), these data are generated with uniform $x \sim U[-1,+1]$, sample size $N=100$, independent Gaussian noise $\varepsilon \sim N(0,1)$, noise level values λ ranging from 0 to 1, as:

Linear: $y = 0.5x + 3\lambda\varepsilon$

W-shaped: $y = |x + 0.5|Ind(x<0) + |x - 0.5|Ind(x \geq 0) + 0.75\lambda\varepsilon$

Sinusoidal: $y = \cos(8\pi x) + 3\lambda\varepsilon$

Circular: $y = z(1-x^2)^{1/2} + 0.9\lambda\varepsilon$, where $z$ is +1 or −1 with equal probability, independent of $x$ and $\varepsilon$.

Figure 1 displays these functions for the noiseless case $\lambda = 0$:

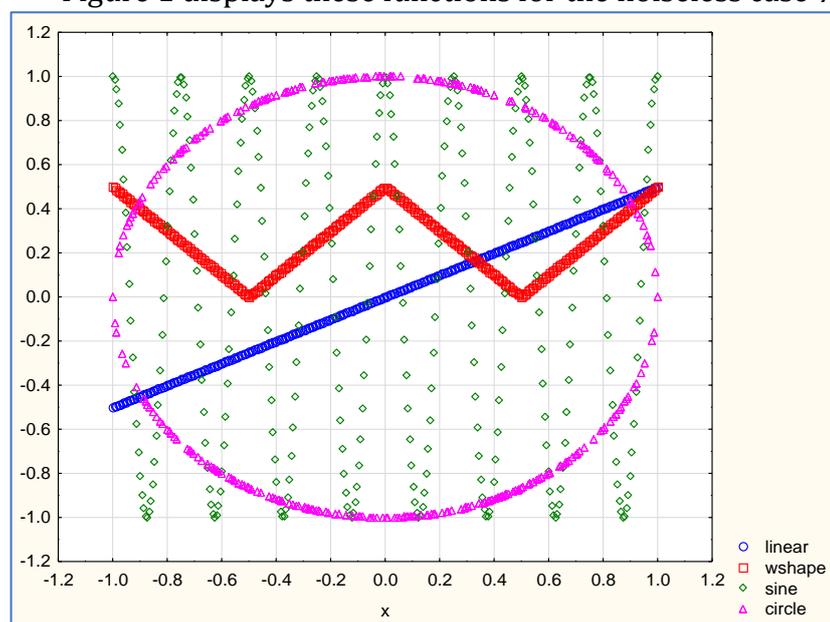

Figure 1: Noiseless toy data for linear, w-shape, sinusoid, circular functions, $N$=300 used in this illustration.

Table 1 provides a list of the notations for the comparison of powers, for different tests and different methods:

Table 1

| | New distance-based Chatterjee | | Original Chatterjee | | Distance correlation |
|---|---|---|---|---|---|
| Tests of independence | H$_0$: dCh[**Y**=**F**(**X**)]=0  dch-PowYfX | H$_0$: dCh[**X**=**G**(**Y**)]=0  dch-PowXfY | H$_0$: Ch[y=f(x)]=0  ch-PowYfX | H$_0$: Ch[x=g(y)]=0  ch-PowXfY | sz-PowYfX |
| Tests of causal direction | H$_0$: Δ(x→y)=0  dch-PowXcY | H$_0$: Δ(y→x)=0  dch-PowYcX | H$_0$: Δ(x→y)=0  ch-PowXcY | H$_0$: Δ(y→x)=0  ch-PowYcX | Not available |

List of the notations for the comparison of powers, for different tests and different methods

Figure 2 displays a comparison of powers for the tests of independence:

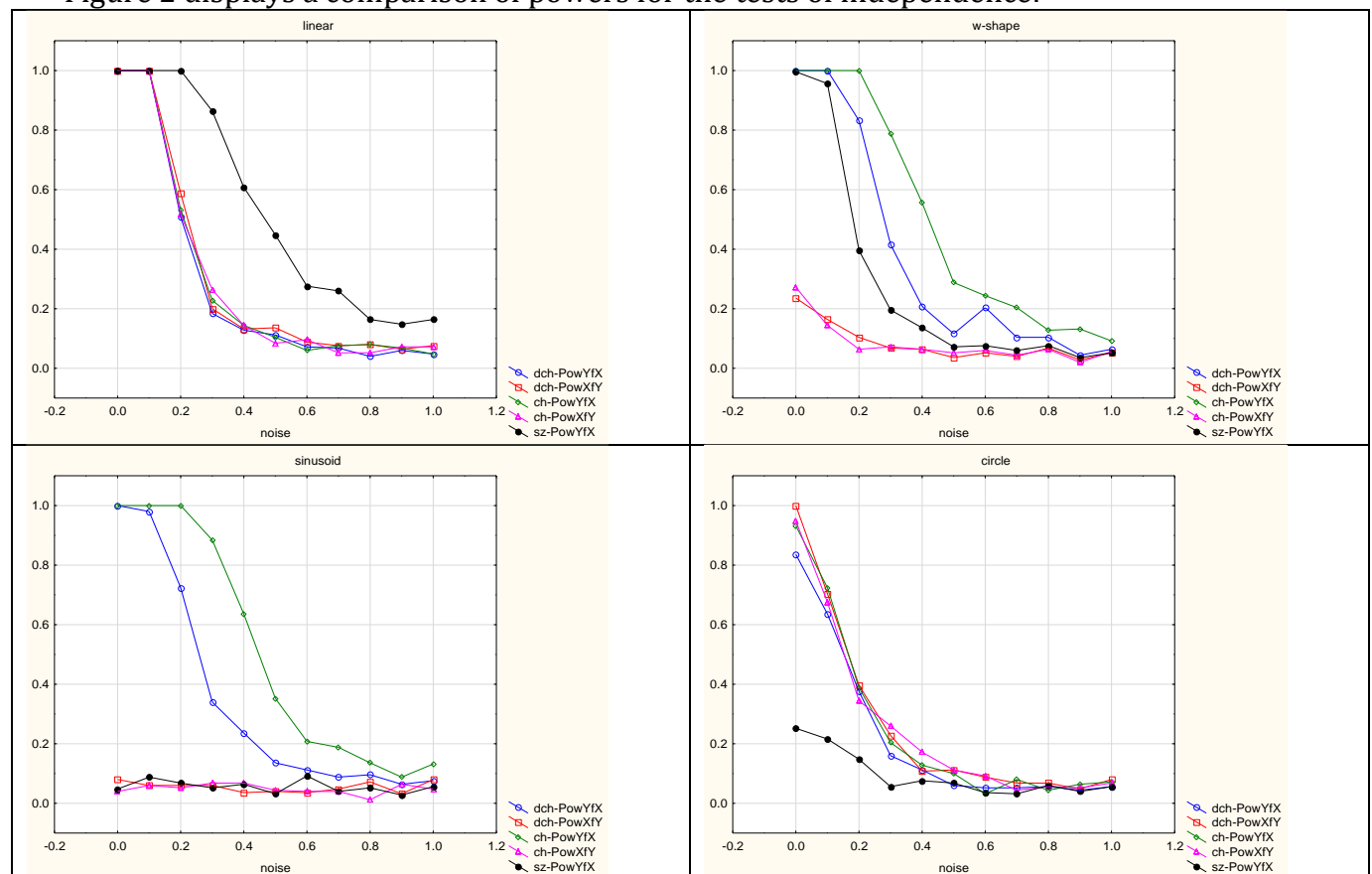

Figure 2: Comparison of powers for the tests of independence, for the two asymmetric functions (*y* predicted by *x*, and *x* predicted by *y*) which are available for the distance-based Chatterjee and original Chatterjee correlations, and for the symmetric distance correlation. Four functions (linear, w-shape, sinusoid, circle) were studied, for sample size *N*=100, at 11 noise levels for λ=0,0.1,0.2,…,1.0; using 250 synchronized random permutations; using 250 repetitions for power computations. Table 1 describes the notations for the different methods.

The results shown in Figure 2 replicate the comparisons reported in Chatterjee (2021) between the original Chatterjee correlation and the distance correlation (Székely et al 2007). The new results here related to power for testing independence show that:
1. Both the distance-based and original Chatterjee correlations have similar power for the linear and circle functions.
2. The distance-based Chatterjee correlation has lower power than the original Chatterjee correlation for the w-shape and sinusoid functions.
3. For the cases of the directed functions w-shape and sinusoid, both the distance-based and original Chatterjee correlation perfectly distinguish that "*y* predicted by *x*" has much higher power than the inverse.

Figure 3 displays a comparison of powers for the tests of causal direction:

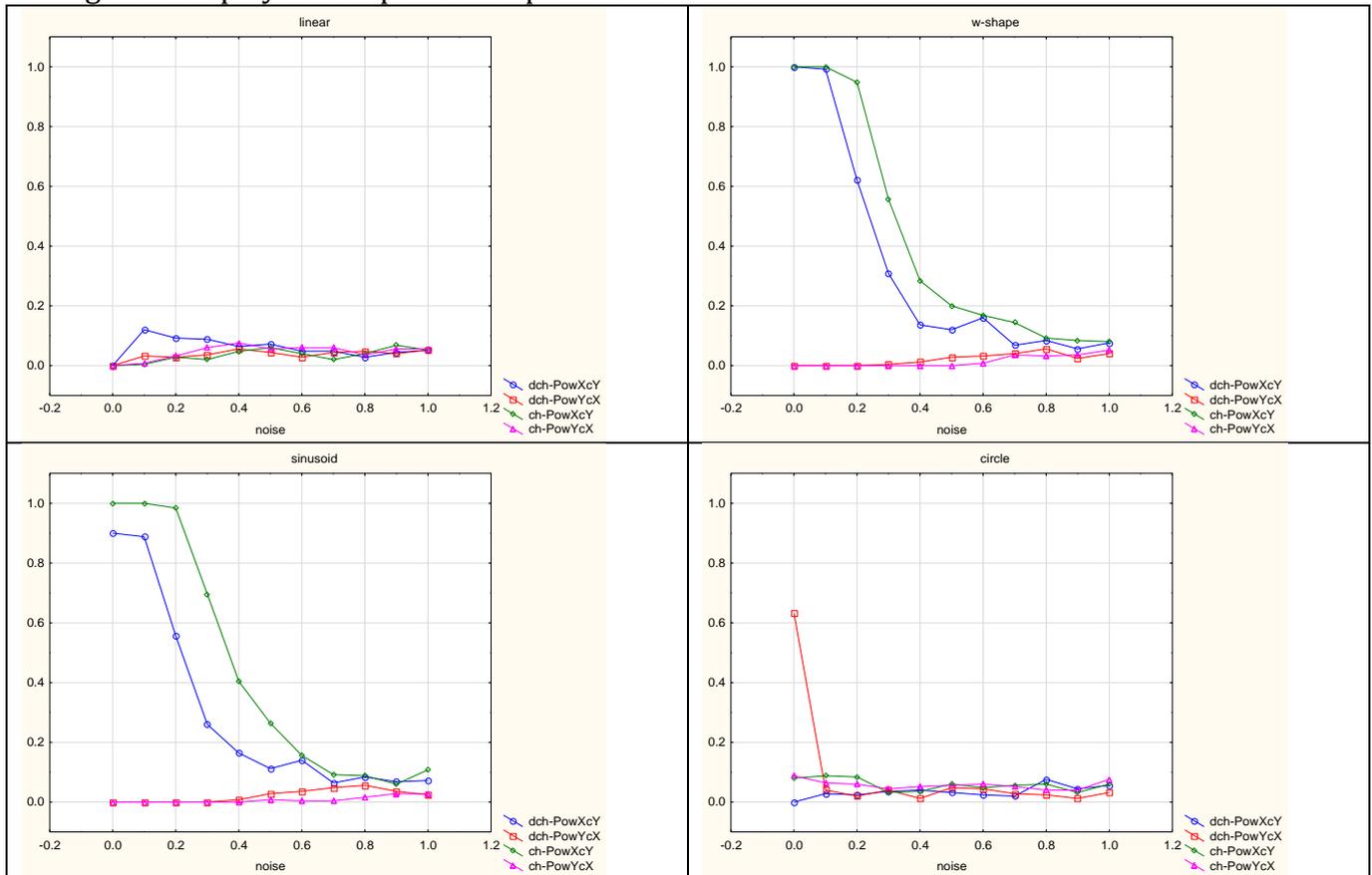

Figure 3: Comparison of powers for the tests of causal direction in the sense of RECI, for the two asymmetric functions (*x* causes *y*, and *y* causes *x*) which are available for the distance-based Chatterjee and original Chatterjee correlations only. Four functions (linear, w-shape, sinusoid, circle) were studied, for sample size *N*=100, at 11 noise levels for λ=0,0.1,0.2,…,1.0; using 250 synchronized random permutations; using 250 repetitions for power computations. Table 1 describes the notations for the different methods.

The results in Figure 3 related to power for testing causal direction in the sense of RECI show that:
1. Both the distance-based and original Chatterjee correlations have near zero power for the linear and circle functions, as expected, due to symmetry of the functions.
2. At low noise levels for the w-shape and sinusoid functions, both the distance-based and original Chatterjee correlation perfectly distinguish with near unit power that "*x* causes *y*", and near zero power for the inverse.
3. At higher noise levels , the distance-based Chatterjee correlation has lower power than the original Chatterjee correlation for the w-shape and sinusoid functions.
4. The distance correlation is a symmetric measure of association and does not distinguish direction of association and does not appear in Figure 3.

## 9. Example for bivariate real-world data

An example is taken from an extensive study in causal inference performed by Mooij et al (2016). It consists of outdoors/indoors temperature measurements, sample size *N*=168. The original data appears in Hipel and McLeod (1994). It is argued in Mooij et al (2016) that outdoor temperatures have a strong impact on indoor temperatures and not the other way around, simply because of the larger outdoors heat capacity. At the moment of this writing, the data is available at:
"http://webdav.tuebingen.mpg.de/cause-effect/"
as "pair0048.txt".

Figure 4 displays the data:

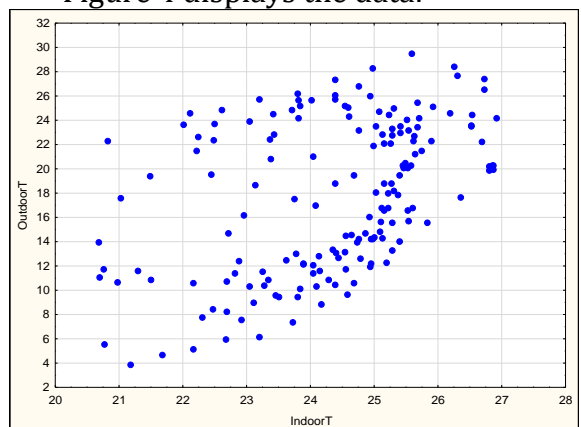

Figure 4: Outdoors/indoors temperature measurements, data from Hipel and McLeod (1994), studied in Mooij et al (2016).

Table 2 shows the "*p*-values", obtained via 300 synchronized random permutations, for the tests of independence and of causal direction, for the new distance-based Chatterjee correlation, the original Chatterjee correlation, and the distance correlation:

Table 2

|  | New distance-based Chatterjee | | Original Chatterjee | | Distance correlation |
|---|---|---|---|---|---|
| Tests of independence | $H_0$: dCh[**Y**=**F**(**X**)]=0 $p$=0.297 | $H_0$: dCh[**X**=**G**(**Y**)]=0 $p$=0.000 | $H_0$: Ch[y=f(x)]=0 $p$=0.033 | $H_0$: Ch[x=g(y)]=0 $p$=0.000 | sz-YfX $p$=0.000 |
| Tests of causal direction | $H_0$: $\Delta$(x→y)=0 $p$=0.880 | $H_0$: $\Delta$(y→x)=0 $p$=0.120 | $H_0$: $\Delta$(x→y)=0 $p$=1.000 | $H_0$: $\Delta$(y→x)=0 $p$=0.000 | Not available |

"*p*-values", obtained via 300 synchronized random permutations, for the tests of independence and of causal direction, for the new distance-based Chatterjee correlation, the original Chatterjee correlation, and the distance correlation:

These results are in favor of significant association between outdoors/indoors temperature. In addition, and in distinction to the distance correlation, both distance-based and original Chatterjee correlations provide evidence for the directed causal influence of outdoor temperature (*y*) on indoor temperature (*x*).

Note: The reason for choosing this one example was that the variables were continuous with no ties for listed values. In most other data files data was not truly continuous, e.g. temperatures were measured with no decimal points, leading to excessive ties. While this situation has a solution available, this present work is restricted to truly continuous data with near-zero probability of ties.

# 10. Example for multivariate simulated data: linear, squares, logarithm of squares

The examples presented here were motivated by the empirical results found in Székely et al (2007), where they use three cases of multivariate associations to compare their method "distance correlation" to other statistics in terms of power analysis. Note that the original Chatterjee correlation will be absent since it is not available for multivariate data.

The data was defined and generated as follows: $\mathbf{X}_i, \mathbf{Y}_i \in \mathbb{R}^{p \times 1}$, with equal dimension $p=5$, with sample size $i=1...N$ ranging from $N=25$ to $N=240$. In the linear and quadratic case, the variables $x_j$ with $j=1...p$ were independent uniformly distributed on [-1,+1]; in the log-quadratic case the variables $x_j$ with $j=1...p$ were independent standard Gaussian. The additive noise variable was independent Gaussian noise $\varepsilon \sim N(0,1)$. The functions were:

- Linear: $y_{ij} = 0.5 x_{ij} - 0.1 \sum_{k \neq j: k=1}^{p} x_{ik} + 0.4 \varepsilon_{ij}$ , $i=1...N$ , $j=1...p$

- Quadratic: $y_{ij} = x_{ij}^2 - 0.1 \sum_{k \neq j: k=1}^{p} x_{ik}^2 + 0.2 \varepsilon_{ij}$ , $i=1...N$ , $j=1...p$

- Log-Quadratic: $y_{ij} = \ln x_{ij}^2$ , $i=1...N$ , $j=1...p$

Figure 5 displays power for the tests of independence and of causal direction:

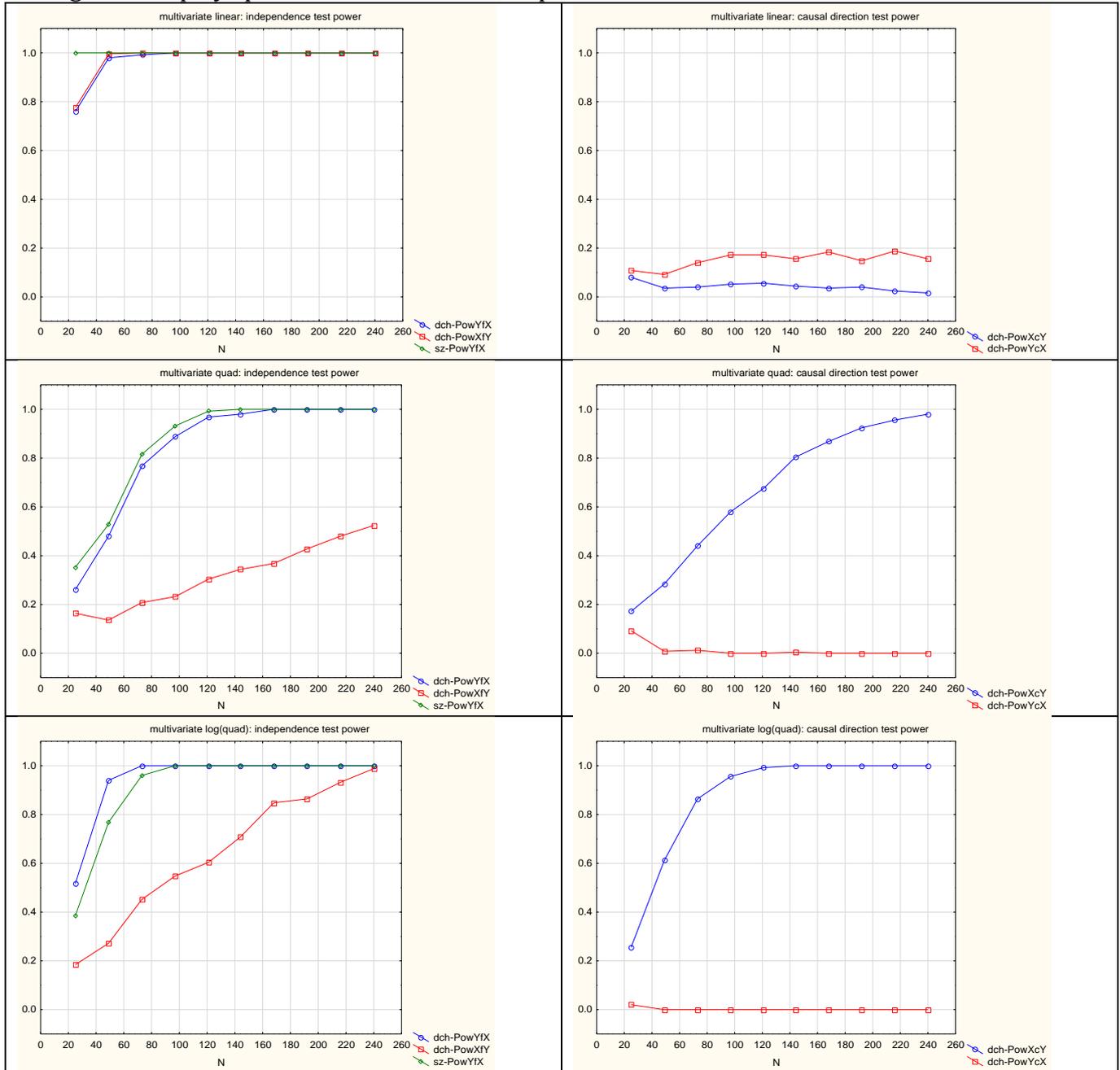

Figure 5: Power curves for the tests of independence (left column) and for tests of causal direction (right column), for three different multivariate functions (linear, quadratic, log-quadratic). The measures of association are distance-based Chatterjee and distance correlation. Table 1 lists the tests and notations. Sample size varied from 25 to 240; and use was made of 250 synchronized random permutations, and 250 repetitions, for power computations. Causal direction statistics are available for the distance-based Chatterjee correlation only.

The results in Figure 5 show that:
1. For the tests of independence, distance correlation has slightly higher power than distance-based Chatterjee correlation only for small sample size.
2. From the tests of independence, the distance-based Chatterjee correlation indicates the causal direction of the multivariate association for the non-linear cases: "X predicts Y, i.e. Y=F(X)" at much higher power that the inverse.
3. As expected, causal direction is not defined for the linear relation.

4. From the tests for causal direction, the correct result using distance-based Chatterjee correlation is obtained for the non-linear associations, especially at relatively large sample size.

## 11. Example for nonlinear complex-valued multivariate simulated data: phase-amplitude coupling

Phase-amplitude coupling, a form of cross-frequency coupling, has been observed across the cortex and appears to be an essential mechanism that mediates information transfer, thus integrating spatially distant functional systems (Canolty and Knight, 2010). In particular, the most common form of coupling occurs between the phase of theta activity (4 to 8 Hz) and the amplitude of gamma activity (>30 Hz) (Canolty and Knight, 2010). Many different statistical methods have been developed with the aim of testing for the presence of phase-amplitude coupling (see e.g. Penny et al 2008).

From a physics and mathematics point of view, phase-amplitude coupling consists of the nonlinear interaction of waves oscillating at different frequencies. This results in higher order interactions, as shown in Kim and Powers (1979). Using equation 18 in Kim and Powers (1979), theta-gamma coupling can be written as:

**Eq. 18**  $y_1(\gamma + \theta) = a_1 x_1(\gamma) x_2(\theta)$

**Eq. 19**  $y_2(\gamma - \theta) = a_2 x_1(\gamma) x_2^*(\theta)$

where all variables are complex valued (e.g. Fourier transform coefficients or analytic signal via the Hilbert transform) including the coupling coefficients $(a_1, a_2)$, and where $x_1(\gamma)$ denotes the gamma oscillation component, which is nonlinearly interacting with $x_2(\theta)$ which is the theta oscillation component, and which combine by multiplication as shown in Eq. 18 and Eq. 19, causally producing the two new oscillation components at frequencies $(\gamma \pm \theta)$. The corresponding time domain representation can be found in Berman et al (2012), their equations 2 and 3 therein.

Based on Eq. 18 and Eq. 19, toy data was generated, as follows for $\mathbf{X}_i, \mathbf{Y}_i \in \mathbb{C}^{2 \times 1}$:

**Eq. 20**  $\begin{cases} y_{i1} = a_1 x_{i1} x_{i2} + \lambda \varepsilon_{i1} \\ y_{i2} = a_2 x_{i1} x_{i2}^* + \lambda \varepsilon_{i2} \end{cases}$

with unit modulus random complex coefficients $a_1$ and $a_2$, $x_{i1}, x_{i2}, \varepsilon_{i1}, \varepsilon_{i2} \sim N_\mathbb{C}(0,1)$ independent standard complex Gaussian, and noise level values λ ranging from 0 to 2.

From the generated data, distance-based Chatterjee correlation and distance correlation were computed on $\mathbf{X}_i, \mathbf{Y}_i \in \mathbb{C}^{2 \times 1}$, without any information about the nonlinear generative mechanism.

Figure 6 displays power for the tests of independence and of causal direction, for this nonlinear complex-valued multivariate example motivated by phase-amplitude coupling:

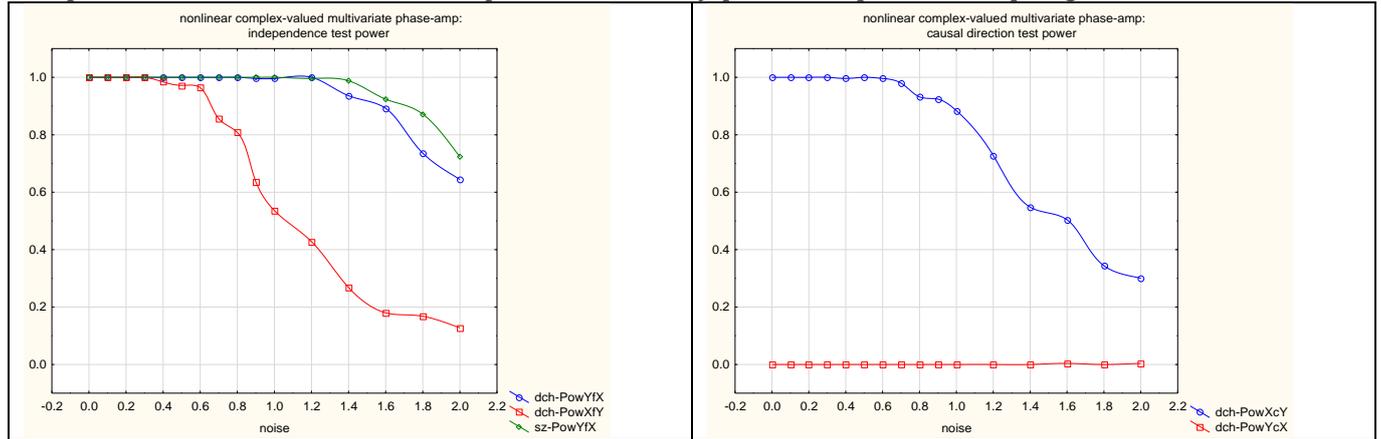

Figure 6: Power curves for the tests of independence (left) and for tests of causal direction (right), for the nonlinear complex-valued multivariate example motivated by phase-amplitude coupling. The measures of association are distance-based Chatterjee and distance correlation. Table 1 lists the tests and notations. Sample size was *N*=100; and use was made of 250 synchronized random permutations, and 250 repetitions, for power computations. Causal direction statistics are available for the distance-based Chatterjee correlation only.

The results in Figure 6 show that:
1. For the tests of independence, distance correlation and distance-based Chatterjee have essentially the same power.
2. From the tests of independence, the distance-based Chatterjee correlation indicates the causal direction of the multivariate association: "X predicts Y, i.e. Y=F(X)" has much higher power that the inverse. This means that the two original interacting frequency components in **X** causally generate the other two new frequency components in **Y**, as expected. The tests for causal direction give the actual confirmation for **X** causing **Y**, especially at relatively low noise contamination.

There are three very important results here:
1. The method presented here is actually a form of directed non-parametric bicoherence (see Kim and Powers 1979, equation 19 therein).
2. The method presented here can be generalized to a multivariate directed non-parametric version, which could include more than just one single pair of interacting waves:

**Eq. 21** $$\begin{cases} y_k = a_k x_k x_{k+1} + \varepsilon_k \\ y_{k+1} = a_{k+1} x_k x_{k+1}^* + \varepsilon_{k+1} \end{cases} \text{ for } k = 1,3,5,\ldots$$

3. In the most general sense, the method presented here is actually a directed non-parametric measure of higher order association, where a response variable can be associated with two or three or more interacting predictors, as in e.g. $y = f\left(x_1 x_2, x_1 x_3, x_2 x_3, x_1 x_2 x_3 \ldots\right)$.

Although not related to the methods presented here, recent work on the application of bispectral analysis techniques for cross-frequency coupling can be found, e.g. in Basti et al (2024).

## 12. Example for outliers in nonlinear multivariate simulated data

The Chatterjee correlation (2021) and the new distance-based Chatterjee correlation are expected to have certain robustness properties against outliers due to the use of ranks. On the other hand, distance correlation (Székely et al 2007) has poor robustness properties (Leyder et al 2024).

Interestingly, Székely and Rizzo (2009 and 2013) proposed the "rank-dCov", with the following prescription according to its verbal description:
1. given two real-valued variables $x$ and $y$;
2. rank them and denote the ranks as $r_x$ and $r_y$;
3. compute the distance correlation of the integer-valued rank variables $r_x$ and $r_y$.

The "rank-dCov" measure is limited to two real variables, it has lower power than the original distance correlation, but is expected to be more robust. Note that according to the prescription for "rank-dCov", it cannot be applied to multivariate data real or complex-valued, since there is no unique way to rank multivariate real or complex valued data.

Note that the rank-dCov differs at a very fundamental level from the distance-based Chatterjee correlation proposed here:
"In the distance-based Chatterjee correlation, the distance transform of the data is ranked and not the data, which guarantees a certain level of robustness and allows for application to real and complex-valued multivariate data, and additionally allows for causal direction inference".

An example described above as the nonlinear multivariate logarithm of squares is re-used here: $\mathbf{X}_i, \mathbf{Y}_i \in \mathbb{R}^{p \times 1}$, with equal dimension $p=5$, with sample size $i=1...N$ ranging from $N=25$ to $N=240$, with independent standard Gaussian $x_j$ with $j=1...p$, and:

- Log-Quadratic: $y_{ij} = \ln x_{ij}^2$ , $i=1...N$ , $j=1...p$

This generates non-outlier data as was previously analyzed above. And from here, in a copy of this data, 5% of the total number of $y_{ij}$ values were randomly chosen and set to a random value between 200 and 210. The two data sets, with and without outliers, were used for the analysis of power, based on 250 synchronized random permutations and 250 repetitions, for independence and for causal direction tests, for the distance-based Chatterjee correlation and for the distance correlation. Figure 7 displays the results for the data with 5% outliers:

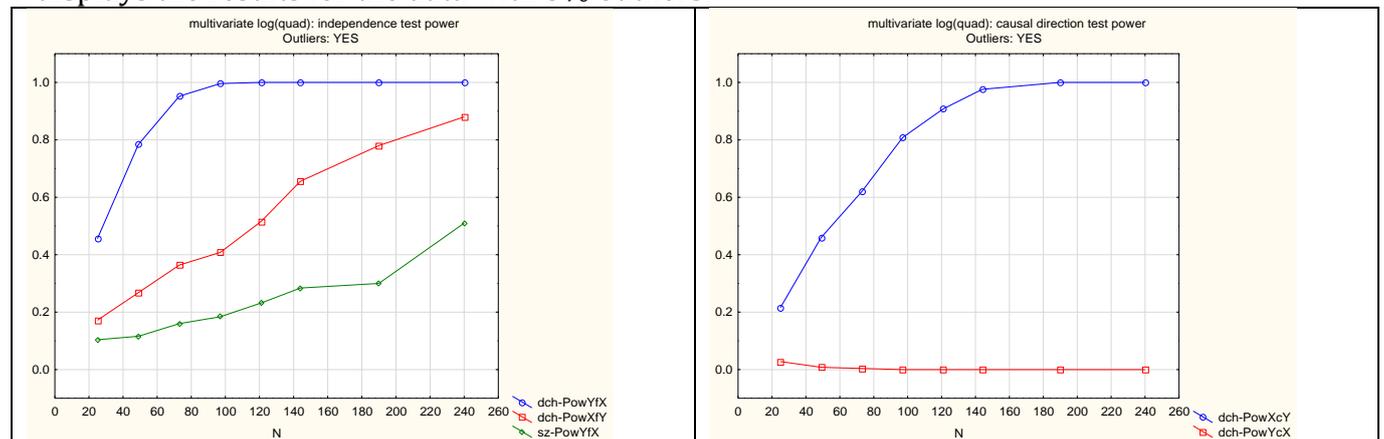

Figure 7: Power curves for the tests of independence (left column) and for tests of causal direction (right column), for multivariate log-quadratic data with outliers. The measures of association are distance-based Chatterjee and distance correlation. Table 1 lists the tests and notations. Sample size varied from 25 to 240; and use was made of 250 synchronized random permutations, and 250 repetitions, for power computations. Causal direction statistics are available for the distance-based Chatterjee correlation only.

The results in Figure 7 show that:
1. For the tests of independence, the distance-based Chatterjee correlation is essentially not affected by outliers (compare to Figure 5, 3rd row left), correctly detecting true association for larger sample sizes, including the correct direction of prediction: "X predicts Y, i.e. Y=F(X)". In contrast distance correlation has very low power in the presence of outliers, and cannot detect true association not even at sample size of N=240.

4. From the tests for causal direction, the distance-based Chatterjee correlation is essentially not affected by outliers (compare to Figure 5, 3rd row right), indicating the correct result in that "X causes Y in the sense of RECI".

## 13. Summary

1. The new distance-based Chatterjee correlation consists of computing the original Chatterjee correlation (2021) on the "distance transformed data". This provides an extension of the original method to multivariate real and complex valued data.

2. Empirical results for bivariate real-valued data show that the distance-based Chatterjee correlation has similar or lower power (test of independence) compared to the original Chatterjee correlation. Both distance-based and original Chatterjee correlations perform better than distance correlation (Székely et al 2007) in general except for linear association.

3. Empirical results for bivariate real-valued data show that both distance-based and original Chatterjee correlations can properly detect causal direction in the sense of RECI.

4. Empirical results for multivariate data (real and complex valued cases) show that the distance-based Chatterjee correlation has similar power (test of independence) as compared to distance correlation. However, only the distance-based Chatterjee correlation can properly detect causal direction in the sense of RECI.

5. Empirical results for multivariate data contaminated by outliers show that the distance-based Chatterjee correlation is hardly affected both in terms of powers for testing independence and causal direction. In contrast distance correlation has very low power in the presence of outliers.

6. Why does it work (the new distance-based Chatterjee correlation)? Informally and intuitively: it takes the distance transform for data as used in distance correlation, but instead of computing its corresponding Pearson correlation, it computes its Chatterjee correlation.

7. Chatterjee (2021 and 2022/2023) reviews the literature and discusses the ideas that attempt to extend the original Chatterjee correlation to multivariate data. All the methods reviewed there are different from the new method proposed here based on the use of the distance transform of the data. At the moment of this writing, there does not appear to be publications on the generalization of the Chatterjee correlation to pairs of real and complex valued vector variables of possibly different dimensions, as achieved here.

8. With respect to the estimation of directed dependence, previous related work limited to two real-valued scalar variables, can be found in Junker et al (2021) and in Griessenberger et al (2022). Their method is based on an asymmetric measure of dependence. And their proposed extension to multivariate data is to compute all pairwise measures, not a genuine multivariate measure as proposed in our present work. In our work, we explicitly make use of the earlier method of Blöbaum et al (2019).

9. The distance-based Chatterjee correlation can be used as a directed non-parametric measure of higher order association. This was shown empirically in the analysis of multivariate non-linear complex-valued data, for phase-amplitude coupling.

10. The new measure can be used in applications such as defining directed nonlinear and nonparametric multivariate versions of cross-frequency coherence and phase synchronization from complex valued data (via Fourier and Hilbert transforms of signals).

11. As with the original Chatterjee, the distance-based version has only one disadvantage: it seems to have less power than several popular tests of independence when the signal is smooth and non-oscillatory for small sample size.

12. For the sake of reproducible research, all delphi-pascal code and executables used for this publication can be found at " https://osf.io/ty6sc/ ". Use was made of the "Embarcadero Free Delphi 11 Community Edition" pascal compiler.